\documentclass[12pt]{article}
\usepackage{caption}
\usepackage{epsfig}
\usepackage{color}
\usepackage{psfrag}
\usepackage{verbatim}
\usepackage{amsopn}
\usepackage{appendix}
\usepackage{dsfont,amsmath,amssymb,color,units,pstricks}
\usepackage{amsfonts}
\usepackage{graphicx}
\usepackage{lscape}
\usepackage{fullpage}
\usepackage{tikz}
\usetikzlibrary{matrix}

\newcommand\be{\begin{equation}}
\newcommand\bea{\begin{eqnarray} \nonumber }
\newcommand\ee{\end{equation}}
\newcommand\eea{\end{eqnarray}}

\begin{document}

\unitlength = 1mm
\title{Instabilities in large economies: aggregate volatility without idiosyncratic shocks}

\author{Julius Bonart$^1$, Jean-Philippe Bouchaud$^1$, Augustin Landier$^{2,1}$, David Thesmar$^{3,1}$}

\maketitle

\noindent\small{$1$: Capital Fund Management, 23 rue de l'Universit\'e, 75007 Paris, France.\\ $2$: TSE, 
Manufacture des Tabacs, 21 All\'ee de Brienne 31000 Toulouse, France\\ $3$: HEC \& CEPR, Dept Economics and Finance, 1 rue de la Libération, 78351 Jouy-en-Josas
Cedex, France.}

\begin{abstract}
We study a dynamical model of interconnected firms which allows for certain market imperfections and frictions, restricted here 
to be myopic price forecasts and slow adjustment of production. 
Whereas the standard rational equilibrium is still formally a stationary solution of the dynamics, we show that this equilibrium becomes 
linearly unstable in a whole region of parameter space. When agents attempt to reach the optimal production target too quickly, coordination breaks down 
and the dynamics becomes chaotic.  In the unstable, ``turbulent'' phase, the aggregate volatility of the total output remains 
substantial even when the amplitude of idiosyncratic shocks goes to zero or when the size of the economy becomes large. In other words, crises become 
endogenous. This suggests an interesting resolution of the ``small shocks, large business cycles'' puzzle.
\end{abstract}

\section{Introduction}

One of the remarkable conundrum in theoretical economics is the so-called ``business cycle'', i.e. the existence of considerable, 
persistent fluctuations of the GDP, even for very large economies, see e.g. \cite{LongPlosser,Cochrane,Bernanke}. 
For example, the quarter-on-quarter growth of the GDP of the US since 1954 has an average of $\approx 3\%$ (annual), 
but with a large rms of $\approx 2.5 \%$ (annual). These fluctuations can culminate in crises, such as the most recent one of 2008. Similar observations  
can also be made on industrial production indices (IPI); for example, the rms of month-on-month IPI growth rate in the US is $\approx 8\%$ (annual) since
1950. 

Naively, however, the output fluctuations of large economies should be very small, because fluctuations in different sectors of the economy should be independent 
and average out. The central limit theorem (CLT) provides a more precise statement: 
for economies made up of $n$ independent sub-sectors of similar sizes, 
the rms of the aggregate output should scale as $1/\sqrt{n}$ and become very small for large $n$. Circumventing this requires either a broad distribution of the 
size of the sub-sectors, or strong correlations between sub-sectors of similar sizes (or a combination of both). The first scenario, advocated by Gabaix
\cite{Gabaix}\footnote{see also, 
in a different context \cite{WB,Farmer,Japs}} appears to be ruled out by the careful empirical study of \cite{Foerster}, who find instead that the correlation between sub-sectors remains large, even at deep 
disaggregated levels. It is plausible that these strong correlations are mediated by the fact that sectors are interconnected, with the input-output network 
providing contagion channels through which small, local output drops can propagate and amplify to become system-wide crises. This scenario was in fact advocated
long ago in a seminal paper by Long \& Plosser \cite{LongPlosser}, followed by a series of papers in the same vein \cite{Horvath,Dupor,Acemoglu}. 
However, the outcome of this strand of research has been somewhat disappointing, in the sense that unless the input-output network has rather special properties (typically a low number $\ll n$ of ``critical'' sectors 
that are major inputs of all other sectors), the aggregate output volatility still behaves as $1/\sqrt{n}$ (times a model dependent prefactor) when 
industry specific shocks are uncorrelated. In other words, unless the whole economy is so ``unbalanced'' that it critically depends on a handful of sectors (like in \cite{Acemoglu2}), 
the origin of the large fluctuations of the aggregate output cannot be rationalized within the existing models of the business cycle. As 
Cochrane puts it \cite{Cochrane}: {\it What shocks are responsible for economic fluctuations? Despite at least two hundred years
in which economists have observed fluctuations in economic activity, we still are not sure.} Although the 2008 crisis can arguably be attributed to the turmoil of the 
financial sector, which is indeed critical to most other sectors of the economy, this scenario is by no means general: the cause of many other substantial activity dips 
cannot be clearly identified. Furthermore, the ``financial sector'' explanation of 2008 only pushes the conundrum one level down: why would such a gigantic 
sector of activity itself be prone to such large shocks?

The aim of this paper is to show that network effects coupled to market imperfections do {\it generically} lead to dynamical instabilities that could be the mechanism
for the large fluctuations. ``Market imperfections'' can mean many different things, such as absence of market clearing due to slow
price adjustments, suboptimal production targets due to frictions, myopic and/or biased expectations (of future prices or future consumptions), etc. We have
actually considered several possible imperfections, and always find that for some parameter values the economy becomes dynamically unstable, in the sense that although 
the (classical) static equilibrium still exists, small fluctuations are amplified and drive the system away from (rather than towards to) this equilibrium. 

Our idea can be schematically understood by considering the following linearized dynamical equation, that describes small fluctuations around the static equilibrium, and
appears in several models including the original Long \& Plosser model \cite{LongPlosser}:
\be\label{lin-dyn}
\vec X_{t+1} = \mathbb{A} \vec X_t +  \vec \varepsilon_t,
\ee
where $\vec X$ describes the set of dynamical variables (i.e. prices, quantities, wages, etc.) and $\vec \varepsilon_t$ represents the idiosyncratic shocks 
(for example, productivity shocks). $\mathbb{A}$ is a dynamical matrix (different from the input-output matrix) that encapsulates all the ingredients of the model -- see below.
Without any market imperfections, the dynamics is found to be stable, in the sense that all the (complex) eigenvalues of $\mathbb{A}$ are of modulus $< 1$ \cite{LongPlosser,Foerster}. 
However, as we shall show below, market imperfections can change the picture completely, and drive one (or more) eigenvalue towards the unit circle. Call $\alpha_+$ the 
eigenvalue of $\mathbb{A}$ with the largest modulus, and $\vec U_+$ its associated eigenvector (such that $||\vec U_+||_2=1$). Suppose (for simplicity) that $\alpha_+$ 
is real and very close to unity: $\alpha_+ = 1 - \eta$ with $\eta \ll 1$. The 
Fourier component $\widehat{\vec X}(\omega)$ of $\vec X_t$ can then be approximated, in the stationary state, as:
\be
\widehat{\vec X}(\omega) \approx \frac{1}{e^{i \omega} - \alpha_+} \hat \epsilon(\omega) \vec U_+ + {\mbox{contribution from other modes}},
\ee
with  $\hat \epsilon(\omega) = \sum_{t=-\infty}^{\infty} e^{- i \omega t}  (\vec \varepsilon_t \cdot \vec U_+)$. Assuming that the idiosyncratic noise $\vec \varepsilon_t$ is 
a white noise of zero mean and variance given by:
\be
\mathbb{E}[\varepsilon_t^j \varepsilon_{t'}^k] = \sigma_j^2 \delta_{jk} \delta_{tt'},
\ee
one can compute in the limit $\eta \to 0$ the correlation function of the components of $\vec X_t$, and find:\footnote{Note that because $||\vec U_+||_2=1$, the 
order of magnitude of $U_+^j \sim n^{-1/2}$.}
\be
\mathbb{E}[X_t^j X_{t}^k] \approx U_+^j U_+^k \, \frac{\Sigma^2}{2 \eta}; \qquad \Sigma^2 = \sum_\ell \sigma_\ell^2 U_+^{\ell2}. 
\ee
This simple result contains some of the important ingredients of our story: it shows that close to an instability, the variance of the fluctuations diverges as $\eta^{-1}$ and 
that there are strong {\it induced correlations} due to the proximity of the instability, since, for $j \neq k$:
\footnote{This scenario is more general and holds whenever $|a_+| \to 1^-$, which will be relevant for our model below.}
\be
\frac{\mathbb{E}[X_t^j X_{t}^k]}{\sqrt{\mathbb{E}[X_t^{j2}]}\sqrt{\mathbb{E}[X_t^{k2}]}} \approx_{\eta \to 0} \frac{U_+^j U_+^k}{|U_+^j| |U_+^k|} = \pm 1
\ee
The last result shows that provided that $X^j$ and $X^k$ are exposed with the same sign to the dominant unstable mode $\vec U_+$, the correlation between 
$X^j$ and $X^k$ tends to unity as the instability is approached, {\it even if their idiosyncratic shocks} $\varepsilon^j, \varepsilon^k$ {\it are completely 
uncorrelated}. These features suggest a promising mechanism to understand the major ``stylized facts'' of the business cycles. 

But why should the economy be close to an instability to start with? As we will find below, the dynamics beyond the (linear) instability point actually remains well behaved 
thanks to stabilizing non-linear terms, absent in the schematic Eq. (\ref{lin-dyn}) above. The analytical description of the dynamics in this ``turbulent'' phase is difficult, 
but one can expect that when only a few modes have become unstable, the above phenomenology remains qualitatively valid -- volatility and correlations are high 
because the dynamics of all firms/sectors is mostly driven by one (or very few) unstable mode(s). This is confirmed by numerical simulations. Furthermore, the 
dynamics in the unstable phase never settles to any equilibrium state even in the absence of any idiosyncratic noise component $\vec \varepsilon_t$. Therefore, 
aggregate volatility in our story is mostly of {\it endogenous origin},\footnote{The idea that a large fraction of the volatility of economic and 
financial systems is of endogenous origin has been advocated for a long time, see e.g. \cite{Keynes, Minsky, Shiller, Black, Summers}, with many recent papers in the econo-physics literature
proposing more explicit scenarii -- see e.g. \cite{Sornette,JStatPhys, Crisis} and refs. therein.} i.e. the result of the non-linear dynamics of a complex network, rather than induced by small 
idiosyncratic shocks that should indeed vanish (or rather, average out) in large economies. We believe that our scenario of ``aggregate volatility
without idiosyncratic shocks'', mediated by instabilities, is extremely generic and could help solve the business cycle puzzle. A similar conclusion has been reached in a very interesting 
recent paper by Mandel et al. \cite{Mandel}, with which our work has clear similarities but also important conceptual and methodological differences. 
In particular, we explicitly model agents' price expectations and analyze how expectation formation, coupled with adjustment costs, can stabilize or destabilize equilibrium.

The outline of this paper is as follows. We introduce our model in Section 2, which is a dynamical generalisation of the standard network of firms with Cobb-Douglas production functions, 
which we supplement with two types of ``market imperfections'': myopic/heuristic price forecasts and slow production adjustments. 
We show that the standard rational equilibrium is always, by construction, a stationary solution of the dynamics. We then study analytically, in Section 3, 
the linear stability of this equilibrium and discover that it is only stable in a certain region of the parameter space. 
When adjustments are slow enough to compensate for myopia, the dynamics is stable and leads to equilibrium. When adjustment is too quick, however, the 
dynamics becomes quasi-periodic or even chaotic. We show numerically that the volatility of the total output remains large, {\it even for small idiosyncratic shocks and large economies}. We end the paper by a discussion of open problems and possible generalisations. 

\section{A dynamical model with slow adjustments and myopic price forecasts}

\subsection{Setting the stage}

The set-up of our model is within the general class of models studied in the literature, where $n$ firms produce goods $i=1,\dots,n$ in quantity $x_t^i$ at prices $p_t^i$ (at time $t$). The 
input-output matrix $w_{ij}$ enters a Cobb-Douglas production function, relating the quantity $x_t^i$ to the amount of labor $\ell_t^i$ and the amounts of goods $\psi^{ij}_t$, $j=1,\dots, n$ used by $i$
through:\footnote{More generally, one could use the so-called constant elasticity to scale (CES) production function, given by:
$$
 x^i = z^i\left[a\left(\frac{\ell^i}{a}\right)^{-r} + (1-a)\sum_j w_{ij} \left(\frac{\psi^{ij}}{(1-a)w_{ij}}\right)^{-r}\right]^{-b/r} \; .
$$
For $r \to 0$, this boils down to the Cobb-Douglas production function, while for $r \to \infty$ one recovers the Leontieff production technology, 
$x^i = z^i \min_{j\in i}\left[\frac{\ell^i}{a}, \frac{\psi^{ij}}{(1-a)w_{ij}}\right]^b$. All the results reported below are qualitatively similar for 
different values of $r$.}
\be
 x_t^i = z_t^i \left(\frac{\ell_t^i}{a}\right)^{ab} \prod_j \left[\frac{\psi_t^{ij}}{(1-a)w_{ij}}\right]^{b(1-a)w_{ij}}\;,
\ee
where $z_t^i$ is the productivity of the firm, $w_{ij}$ describes the share of input $j$ in the production of $i$, with $\sum_j w_{ij}=1, \forall i$, 
$a \in [0,1]$ is a parameter describing the share of labor in production, and $b$ a parameter describing the dependence of production on overall scale. $b=1$ 
corresponds to constant return to scale (CRS), while $b < 1$ corresponds to decreasing return to scale (DRS). It is customary to assume that $a$ and $b$ are independent of the firm $i$, although this could easily be changed. 
Typically, $a \approx 0.5$ and $b \approx 0.9$, values that we will use in the following.

The households have uniform log-utilities for all goods, which simply means that they consume each good inversely proportionnaly to its price, and spend all their available revenues, made of 
their wages and the dividends coming from the profits of the firms (if these profits are negative, households finance the losses). The labor market and all the goods markets clear, in the sense that
the wage $h_t$ (assumed to be the same for all firms) and the prices $p_t^i$ adjust instantaneously so that supply (of work and goods) equal demand. The total supply of labor is constant in time, and
normalized to unity:
\be
\sum_{i=1}^n \ell_t^i \equiv 1, \qquad \forall t.
\ee

At time $t$,  firm $i$ must decide on its production target for the next time step $t+1$. It observes the current wage level $h_t$ and prices $\{p_t^j\}$, and makes projections for the price 
$\mathbb{E}_t(p_{t+1}^i)$ 
at which it will be able to sell its product at time $t+1$. We assume the firm knows its current productivity level $z_{t}^i$. With these informations, the optimal production level $x_{t+1}^{i*}$ comes from maximising 
the discounted expected profits minus costs, ${\cal P}_t^i$:
\be
{\cal P}_t^i \equiv \beta_t \hat x_{t+1}^{i} \mathbb{E}_t(p_{t+1}^i) - h_t \ell_t^i - \sum_{j=1}^n p_t^j \psi_t^{ij},
\ee
under the constraint:
\be\label{constraint_1}
\hat x_{t+1}^{i} = z_{t}^i \left(\frac{\ell_t^i}{a}\right)^{ab} \prod_j \left[\frac{\psi_t^{ij}}{(1-a)w_{ij}}\right]^{b(1-a)w_{ij}}.
\ee
Note that the discount rate $\beta$ may depend on time (see below) but for simplicity we assume it is independent of $i$. When $b < 1$, the above optimisation program has a unique solution, given by:
\be
x_{t+1}^{i*} = \left[ z_{t}^i \left(\beta_t \mathbb{E}_t(p_{t+1}^i)\right)^b h_t^{-ab} \prod_j (p_t^j)^{b(1-a)w_{ij}} \right]^{\frac{1}{1-b}},
\ee
where we have absorbed a factor $b^b$ in a redefinition of $z_t^i$.

\subsection{Slow adjustments}

Now, we depart from the usual assumption that firms are strictly profit maximizers and introduce the idea that the production level cannot change arbitrarily fast from one period to the next. 
This can be due to all sorts of ``adjustment costs'' (difficulty to hire/fire fast enough, or to buy the necessary machines, etc.)\footnote{
The production update rule Eq. (\ref{prod-update}) can actually be seen as resulting from adjustment costs proportional to $(1-\gamma)/\gamma \times 
(x_{t+1}^{i} - x_t^i)^2$.}, but also to a precautionary ``rule of thumb'' that takes into account the 
risk of mis-estimating future prices and productivities (this is sometimes called ``conservatism bias'' \cite{Ward}). It is thus reasonable to assume that the real production target $x_{t+1}^{i}$ of the firm is an average between the current production level and 
the above optimal level, i.e.:
\be\label{prod-update}
x_{t+1}^{i} = (1 - \gamma) x_t^i + \gamma x_{t+1}^{i*},
\ee
where $\gamma$ is a friction parameter, which is small if adjustment costs/risk aversion are large, and close to unity in the opposite case.
Now the firm has to determine how much labor and goods it needs to 
achieve this production level, for the lowest costs. Introducing a Lagrange parameter $\lambda_t^i$, it is easy to find that these quantities are given by:
\be\label{quants}
\ell_i^t = \frac{ab \lambda_t^i  x_{t+1}^{i}}{h_t}; \qquad \psi_t^{ij} = \frac{(1-a)b w_{ij} \lambda_t^i  x_{t+1}^{i}}{p_t^j};
\ee
where $\lambda_t^i$ is fixed such that Eq. (\ref{constraint_1}) is satisfied with $\hat x_{t+1}^{i}=  x_{t+1}^{i}$. This leads to:
\be\label{lagrange}
\lambda_t^i = \beta_t \mathbb{E}_t(p_{t+1}^i) \left[\frac{x_{t+1}^{i}}{x_{t+1}^{i*}}\right]^{\frac{1-b}{b}}.
\ee
Note that when $\gamma=1$ (no friction), $\lambda_t^i = \beta_t \mathbb{E}_t(p_{t+1}^i)$. Note that our main result below 
(that the economy is unstable when expectations are not rational) holds in an economy where $\gamma=1$, i.e. in the absence of adjustment costs. 
As a matter of fact, adjustment costs turn out to be crucial to {\it recover} (in some regimes) the general equilibrium situation even 
in the absence of rationality!

\subsection{Market clearing conditions}

Using the assumption that the labor market clears immediately gives the wage at time $t$, since:\footnote{As discussed below, this is not 
entirely consistent with the assumption that firms know the wage before deciding their production target. We do not attempt to describe in 
detail who the labor market clears, but just assume it does.}
\be\label{wages}
\sum_{i=1}^n \ell_t^i = 1 \longrightarrow h_t = ab \sum_{i=1}^n \lambda_t^i  x_{t+1}^{i}.
\ee
Clearing of the good markets is slightly more tricky and requires a discussion of possible time lag effects. A natural assumption would that the wealths 
$M_t$ available to the households at time $t$ come from the wages and dividends on profits at time $t-1$, i.e.:
\be
M_t = \underbrace{h_{t-1}}_{\mbox{wages}} + \underbrace{\left[\sum_{k=1}^n x_{t-1}^k p_{t-1}^k - h_{t-1} - \sum_{k=1}^n \sum_{j=1}^n \psi_{t-1}^{kj} p_{t-1}^j \right]}_
{\mbox{profits/losses}}.
\ee
However, this makes the model slightly more complex as it introduces an extra time lag and requires the introduction of an interest rate. 
In order to keep the setting of the model and the algebra as simple as possible, we choose instead to 
model all payment and consumption processes as instantaneous. In other words, at time $t$ many things happen ``quickly'': wages are paid to household, 
firms buy the input goods and make profits that are also paid to households, who consume immediately the goods produced at $t$, the prices of which adapt such that
markets clear. This is of course slightly absurd, but introducing an extra time lag does not change the phenomenology of the model, only the precise value of the 
parameters where the instability sets in. Therefore, we write:
\be
M_t = \sum_{k=1}^n x_{t}^k p_{t}^k  - \sum_{k=1}^n \sum_{j=1}^n \psi_{t}^{kj} p_{t}^j =  \sum_{k=1}^n x_{t}^k p_{t}^k  - (1-a)b  \sum_{k=1}^n \lambda_t^k  x_{t+1}^{k},
\ee
where we have used Eq. (\ref{quants}) for $\psi_{t}^{kj}$ and $\sum_j w_{kj}=1$. Market clearing for product $i$ at time $t$ then reads:
\be
x_t^i = \frac{M_t}{n p_t^i}  + \sum_{j=1}^n \psi_t^{ji},
\ee
where the first term is the demand from households and the second term is the demand from other firms. The market clearing conditions finally read:
\be\label{market-clearing}
x_t^i p_t^i - \frac1n \sum_{k=1}^n x_{t}^k p_{t}^k = (1-a)b  \sum_{j=1}^n \left(w_{ji} - \frac1n \right) \lambda_t^j  x_{t+1}^{j} 
\ee
Note that when $\gamma=1$ this forward-looking equation has a simple property that pre-announces the instabilities that we will find below. As noted above, for $\gamma=1$ one has  
$\lambda_t^i = \beta_t \mathbb{E}_t(p_{t+1}^i)$. Assuming price forecasts are un-biased, i.e. $p_{t+1}^i = \mathbb{E}_t(p_{t+1}^i)$+ noise\footnote{
We assume that the noise term is not correlated with any past information (e.g. the $x_t^i$) up to $t+1$, as customary in rational equilibrium theory.}, and introducing the
vector $S_t^i = x_t^i p_t^i$, one immediately sees that the dynamics of the vector $\vec S_t^\perp$ in the subspace orthogonal to the uniform vector $\vec 1$ writes:
\be \label{TR-cond}
\vec S_{t+1}^\perp =  \frac{1}{\beta_t(1-a)b} [\mathbb{W}^T]^{-1}\vec S_{t}^\perp + {\mbox {noise}}.
\ee
But since all the eigenvalues of $\mathbb{W}^T$ are of modulus $< 1$, and the product $\beta_t(1-a)b$ is itself $< 1$, one sees that the above iteration is always exponentially unstable,
unless $S_{t}^\perp \equiv 0$ (in which case the market clearing condition is identically satisfied). This is called the {\it transversality condition}, which is obeyed when 
agents optimize their inter-temporal utility function, as in the Long-Plosser model discussed below (see section \ref{LP}). In the general case however this 
condition does not hold, and we will find that the dynamics is only stable if adaptation is slow enough, i.e. when $\gamma$ is smaller than a certain value
$\gamma_c$ that we will compute below.

\subsection{Expected price: extrapolative, myopic or mean-reverting rules}

We are now in the position to ``close'' the model and write down dynamical equations for the deviations from equilibrium. In order to do this, we need to specify how 
the expected future discounted price $\beta_t \mathbb{E}_t(p_{t+1}^i)$ is determined. For the price, we posit that firms have 
``extrapolative expectations'', in the sense that:
\be \label{price-update}
\mathbb{E}_t(p_{t+1}^i) = p_t^i \left(\frac{p_t^i}{p_{t-1}^i}\right)^q \approx p_t^i  + q (p_t^i - p_{t-1}^i), \qquad q \in [-1,1] 
\ee 
which means that firms assume the future price is the current price, 
plus a correction related to the recent trend on the price, which is small when $|p_t^i - p_{t-1}^i| \ll p_t^i$. 
When $q > 0$, firms expect the recent trend to persist, while when $q < 0$, they assume some mean reversion will take place. When $q=0$, the 
expected future price is simply the current price, and when $q=-1$, the future price is expected to be given by the last price. 
Along the same line of thought, it is reasonable to assume that the discount rate 
$\beta_t$ is related to the latest inflation indicator, i.e.:
\be
\beta_t =  \beta_0  \left[\left(\prod_{i=1}^n \frac{p_t^i}{p_{t-1}^i}\right)^{\frac{1}{n}}\right]^{-q_0}.
\ee
In other words, if prices are expected to rise on average between $t$ and $t+1$, the discount factor $\beta_t$ should be less than unity. The coefficient 
$\beta_0$ can always be set to unity up to a multiplicative shift of the productivities $z^i$. The natural choice is $q_0=q$ (meaning that 
firms adapt their price and the global price level consistently), although other possibilities 
can be considered as well. With these last ingredients, the dynamics of the system is fully specified. Note that our dynamical equations obey a ``monetary unit symmetry'' (MUS), i.e. they are unchanged if all prices and wages are multiplied by an arbitrary constant, as it should be. 

\subsection{Summary} 

Before moving on to analyze the equilibrium and its stability, it might be useful to give a synthetic recap of the logic of our model. At time $t$, firms 
decide on the quantity they want to produce at the next time step. In order to do this, they need an estimate of the price $\mathbb{E}_t(p_{t+1})$ 
at which they will be able to sell their products at time $t+1$. This they do by using past prices and the simple rule, Eq. (\ref{price-update}). Once this 
price is known, they compute the optimal quantity $x_{t+1}^*$ that maximizes expected profits, with a known Cobb-Douglas technology. Firms actually 
decide not to produce $x_{t+1}^*$ but to make a fraction $\gamma$ of the distance between the current production $x_t$ and the optimal production $x_{t+1}^*$.
Knowing this ``compromise'' target production, they can now decide on the optimal amount of labor and inputs, that minimize the production costs, knowing the current prices and 
wages. This leads to Eqs. (\ref{quants}) \& (\ref{lagrange}). Finally, all at once at time $t$, firms sell the production they decided at $t-1$, pay wages \& dividends, and buy the inputs for the
next production, while households buy firms production, and prices at time $t$ are such that markets clear. This set of rules are enough to fully 
specify the dynamics of the model. Many simplifying assumptions can be questioned, such as for example the simultaneity of the money flows and the fact 
that markets clear instantaneously. However, by keeping the framework as simple as possible, we will be able to show that there is a generic transition 
line between a stable regime where the standard rational equilibrium is reached, and an unstable regime where chaotic dynamics sets in, leading to 
endogenous volatility. As we will mention in the final section, these conclusions appear to be robust against many of the above simplifying assumptions (see 
also \cite{Mandel} for similar conclusions).

\section{Equilibrium and linearized dynamics}

\subsection{The equilibrium conditions}

If productivities are fixed in time, i.e. $z_{t}^i \equiv {\overline{z^i}}$, a 
static equilibrium exists such as $p_t^i = p_{eq}^i$, $x_t^i = x_{eq}^i$ and $\lambda_t^i = \lambda_{eq}^i = \beta_0 p_{eq}^i$. Clearly, from Eq. (\ref{prod-update}), the 
equilibrium production coincides with the optimal one, $x_{eq}^i = x^{i*}$ with, from Eq. (\ref{price-update}), $\mathbb{E}(p^i) \equiv p_{eq}^i$. Since there is
no inflation, $\beta=\beta_0 \equiv 1$. This leads to the following standard equilibrium relations that set prices, productions and wage: 
\be\label{equilibrium}
\vec V_{eq} - \frac{\vec V_{eq} \cdot \vec 1}{n} \vec 1 = (1-a) b \widehat{\mathbb{W}}\, \vec V_{eq}; \qquad h_{eq} = ab (\vec V_{eq} \cdot \vec 1),
\ee
with $(\vec V)_{eq}^i \equiv x_{eq}^i p_{eq}^i$ (called -- up to a normalisation -- the ``influence vector'' in \cite{Acemoglu}), $(\vec 1)^i \equiv 1$, $\widehat{\mathbb{W}}_{ij} = w_{ji} - \frac1n$, and:
\be
x_{eq}^{i} = \left[ {\overline{z^i}} \left(p_{eq}^i \right)^b h_{eq}^{-ab} \prod_j (p_{eq}^j)^{b(1-a)w_{ij}} \right]^{\frac{1}{1-b}}.
\ee

The question is to know whether this equilibrium can ever be reached dynamically, or if any small amount of noise drives the system away from equilibrium, which would make the
whole analysis of the equilibrium situation irrelevant to understand the fluctuations of the aggregate output. What we will find is that generically, there exists a line in the plane $(q,\gamma)$ 
below which the equilibrium is stable, and above which it becomes unstable. In the latter case, the aggregate output volatility is self-induced by the non-linear dynamics of the system, and not related to any 
exogenous ``shocks''. 

\subsection{The linearized dynamical equations}

In order to access the stability of the equilibrium situation, we study the dynamics of small perturbations around equilibrium. We therefore set:
\be
p_t^i \equiv p_{eq}^i \exp({\pi_t^i}); \qquad x_t^i \equiv x_{eq}^i \exp({\xi_t^i});\qquad  \lambda_t^i \equiv  \beta_0  p_{eq}^i \exp({\mu_t^i}); \qquad z_t^i = \overline{z^i} \exp({\epsilon_t^i}),
\ee
with $\pi, \xi, \mu, \epsilon \ll 1$. Expanding the above equations to first order in these quantities leads to the following set of equations:
\begin{align}
(\mathbb{I}-a\mathbb{J}_1) \vec\mu_t &= \left(\frac{1-b}{b}\mathbb{I} + a\mathbb{J}_1\right)\vec\xi_{t+1} + (1-a)\mathbb{W} \, \vec\pi_t - \frac{1}{b}\vec\epsilon_t \;,\label{eq:evolution1}\\
(1-\gamma) (\vec \xi_{t+1}-\vec\xi_t) &=  \gamma\frac{b}{1-b}(\vec\pi_t-\vec\mu_t) -  \gamma \frac{b}{1-b}(q\mathbb{I}- q_0\mathbb{J}_0)
(\vec\pi_{t-1}-\vec\pi_t)  \;, \label{eq:evolution2}\\
(\mathbb{I}-\mathbb{J}_2)(\vec\xi_t + \vec\pi_t) &= (1-a)b (\widetilde{\mathbb{W}}-\mathbb{J}_2) (\vec\mu_t + \vec\xi_{t+1})\;\label{eq:evolution3}.
\end{align}
with the following definition for the five matrices:
\begin{align}
  \mathbb{W}_{ij} &= w_{ij}\;,\\
  \widetilde{\mathbb{W}}_{ij} &=  w_{ji}\frac{V_{eq}^j}{V_{eq}^i}\;,\\
  \mathbb{J}_{0ij} &= \frac{1}{n} \;,\\
  \mathbb{J}_{1ij} &= \frac{V_{eq}^j}{\sum_k V_{eq}^k} \;,\\
  \mathbb{J}_{2ij} &= \frac{V_{eq}^j}{n\, V_{eq}^i}\;.  
\end{align}
where $\mathbb{J}_{0,1,2}$ are projectors with $\mathbb{J}_{1}\times \mathbb{J}_{2}=\mathbb{J}_{1}$, $\mathbb{J}_{2}\times \mathbb{J}_{1}=\mathbb{J}_{2}$,
$\mathbb{J}_{1}\times \widetilde{\mathbb{W}}=\mathbb{J}_{1}$, and $\mathbb{J}_{2}\times \widetilde{\mathbb{W}}=\widetilde{\mathbb{W}}$.
Note that the MUS imposes that whenever $\xi_t \equiv 0$ and $\mu_t = \pi_t \equiv \pi_0$, the linearized dynamical equations should be identically obeyed. 
Using the equilibrium condition Eq. (\ref{equilibrium}) above, one can check that this indeed holds true. 

Note finally that had we kept the more natural one-time 
lag rule between wages \& dividend payments and consumption, only the last equation above would change and would read (for zero interest rate):
\be
\vec\xi_t + \vec\pi_t - (1-a)b \widetilde{\mathbb{W}}\, (\vec\mu_t + \vec\xi_{t+1})= \mathbb{J}_2 \left[(\vec\xi_{t-1} + \vec\pi_{t-1}) - (1-a) b(\vec\mu_{t-1} + \vec\xi_{t})\right]\;\label{eq:evolution3-lag}
\ee

\section{From stable economies to crises prone economies}

\subsection{The Long-Plosser equation}
\label{LP}
The above framework generalizes previous attempts to write dynamical equations for the output and prices in network economies. Let us discuss in particular 
how the Long-Plosser model can be recovered. In the fully rational Long-Plosser model, agents forecast the future prices perfectly and produce optimal quantities, 
which means in the present context that $\lambda_t^i \equiv \beta_0 \mathbb{E}_t[p_{t+1}^i]$, and $x_{t+1}^{i} \equiv x_{t+1}^{i*}$. Inserting the corresponding condition
$\vec \mu_t = \vec \pi_{t+1} +$ noise in Eq. (\ref{eq:evolution3}) leads to:
\be
(\mathbb{I}-\mathbb{J}_2)  \vec g_t = (1-a)b (\widetilde{\mathbb{W}}-\mathbb{J}_2) \vec g_{t+1}, \qquad \vec g_t := \vec\xi_t + \vec\pi_t.
\ee
However, since the singular values of $\widetilde{\mathbb{W}}$ are all $< 1$, this forward-in-time iteration is generically {\it unstable}, even more so because of the 
prefactor $(1-a)b$. The {\it only} ``stable path'' of the economy chosen by rational agents is therefore such that $S_t^i=x_t^i p_t^i =$ constant for all $i,t$, i.e.
$\vec\xi_t = g_0 \vec 1 - \vec\pi_t$, where $g_0$ is an arbitrary constant 
(transversality condition): prices and quantities are always inversely proportional to one another, 
as indeed found in the Long-Plosser model. Plugging this into Eq. (\ref{eq:evolution1}) and using $\mathbb{W}\vec 1=\vec 1$ yields the Long-Plosser dynamical equation\footnote{Note that Eq. (\ref{eq:evolution2}) 
has no counterpart in the Long-Plosser framework, since the transversality condition completely fixes the dynamics of the quantities $x$.} (see also \cite{Horvath,Dupor}):
\be
\vec\xi_{t+1} =  b(1-a)\mathbb{W} \, \vec\xi_t   + \vec\epsilon_t.
\ee
Since all the singular values of $\mathbb{W}$ are less than unity, and $b(1-a) < 1$, this equation leads, within the one dimensional subspace $\vec S \parallel \vec 1$, to {\it stable 
fluctuations} (compare with Eq. (\ref{TR-cond}), which leads -- for the very same reasons -- to unstable dynamics in the subspace $\vec S \perp \vec 1$). The volatility of the total output furthermore tends to zero for large 
economies unless the input-output matrix $\mathbb{W}$ has a very particular star-like structure \cite{Acemoglu}. Actually, the Acemoglu-Carvalho model corresponds to
an idiosyncratic noise $\vec\epsilon_t$ that vary so slowly in time that equilibrium can be reached before the noise has significantly changed. In this ``adiabatic'' limit 
(to use an expression from physics to describe slowly changing external conditions), the economy goes through a sequence of quasi-equilibrium situations 
characterized by:
\be
\vec \xi = \left[\mathbb{I} -  b(1-a)\mathbb{W}\right]^{-1} \vec\epsilon
\ee
which is precisely the equation considered in Acemoglu et al. \cite{Acemoglu}. Defining the relative fluctuations of output as a flat average $n^{-1}\sum_i \xi^i$, and using
the definition of the ``influence vector'' $\vec V_{eq}\equiv n^{-1}\vec 1^T\cdot [\mathbb{I} -  b(1-a)\mathbb{W}]^{-1}$, 
one obtains the volatility of aggregate production as:
\begin{equation}
\Sigma^2_{slow} = \sum_{\ell=1}^n \sigma_\ell^2 \vec V_{eq}^{\ell \,\, 2} \quad \leq \quad
\Sigma^2_{fast} = n^{-2} \sum_{i,j=1}^n \sum_{k=1}^n {\cal M}^{ij,kk} \sigma_k^2,
\end{equation}
with
\begin{equation}
({\cal M}^{-1})^{ij,k\ell} = \delta_{ik}\delta_{j\ell}-b^2(1-a)^2 \mathbb{W}^{ik} \mathbb{W}^{j \ell}.
\end{equation}
The first (``slow'') result holds in the slow adiabatic limit of \cite{Acemoglu}, where the shocks are essentially permanent on the time scale needed to
reach equilibrium, while the second (``fast'') result holds when $\vec \epsilon_t$ is a quickly evolving white noise, with 
the assumption that shocks are idiosyncratic and with the same variance in both cases (i.e. $\mathbb{E}(\epsilon^i \epsilon^j) = \sigma_i^2 \delta_{ij}$).\footnote{The intermediate 
case when $\vec \epsilon_t$ has non trivial temporal correlations can also be treated by going in Fourier space. However, the final result is not very 
telling.}  
However, as discussed in \cite{Foerster} and emphasized in the introduction above, this family of stable dynamical equation cannot explain large cross-correlations between sectors
when shocks are idiosyncratic. The empirical input-output matrix is not ``star-like'' enough to prevent $\Sigma^2$ from being much too small at large $n$.

The whole idea of our framework is to relax the very restrictive assumptions of Long-Plosser (and subsequent papers), 
whereby agents perfectly predict the future and economies necessarily follow a stable path from now to infinite times.
The general equations obtained above only assume an imperfect and myopic optimisation scheme, together with a heuristic forecast of future prices. 
As we show now, this can induce dynamical instabilities and a much richer phenomenology, including large volatilities and crises.  

\subsection{The general case: linear stability analysis}
\label{sect_stability}

The stability analysis of Eqs. (\ref{eq:evolution1}, \ref{eq:evolution2}, \ref{eq:evolution3}) in the case of a general stochastic matrix $\mathbb{W}$ is difficult. 
However, the situation simplifies considerably -- without changing the main qualitative conclusions -- when $\mathbb{W}$ is {\it normal}, i.e. when it commutes with its transpose. 
In this case, it is easy to check that $\vec V_{eq} \propto 
\vec 1$, i.e. the equilibrium share $S^i_{eq}$ of firm $i$ in the economy, defined as:
\be
S^i_{eq}= \frac{x_{eq}^i p_{eq}^i}{\sum_k  x_{eq}^k p_{eq}^k}
\ee
is the same for all $i$: $S^i_{eq} = 1/n$. 
One can then decompose the fluctuations $\vec \pi, \vec \xi, \vec \mu$ in the eigenbasis of $\mathbb{W}$, and study each component independently, 
since in this case $\mathbb{J}_{0}=\mathbb{J}_{1}=\mathbb{J}_{2}= 
\vec 1^T \, \vec 1/n$ and $\widetilde{\mathbb{W}}={\mathbb{W}}^T$. 

\subsubsection{The uniform mode}

Let us start with the uniform mode $\vec \pi=\pi \vec 1, \vec \xi=\xi \vec 1, \vec \mu=\mu \vec 1$, which corresponds to the eigenvalue $s = 1$ of $\mathbb{W}$. The linear equations then become:
\begin{align}
(1-a)(\mu_t - \pi_t) &= \left(\frac{1-b}{b} + a\right) \xi_{t+1} - \frac{1}{b} \epsilon_{1,t} \;,\label{eq:evolution1bis}\\
(1-\gamma) (\xi_{t+1}-\xi_t) &=  \gamma\frac{b}{1-b}(\pi_t-\mu_t) -  \gamma \frac{b}{1-b}(q-q_0)(\pi_{t-1}-\pi_t)  \;, \label{eq:evolution2bis}\\
\end{align}
where $\epsilon_{1,t}= \vec \epsilon_t \cdot \vec 1$. Eq. (\ref{eq:evolution3}) turns out to be trivially satisfied, leaving the evolution of $\pi_t$ undermined. 
This means that in the model where payment and consumption are simultaneous, the evolution of the overall price level is undetermined. This is not the case 
when a finite time lag is introduced, such as in Eq. (\ref{eq:evolution3-lag}). Still, when $q=q_0$, the evolution of the overall price level is, as expected, 
totally irrelevant and we will for simplicity focus on this case here, commenting on more general cases below.

The combination of Eqs. (\ref{eq:evolution1bis},\ref{eq:evolution2bis}) leads, for $q=q_0$ to:\footnote{When $q \neq q_0$, and extra term $(q-q_0)(\pi_t - \pi_{t-1})$
appears in the right hand side of the equation, which would not affect the stability analysis reported below.}
\begin{equation}
(1 - \gamma + \zeta (1-b+ab)) \xi_{t+1} = (1-\gamma) \xi_{t} + \zeta \epsilon_{1,t}; \qquad \zeta = \frac{\gamma}{(1-a)(1-b)}.
\end{equation}
Since $\zeta (1-b+ab) \geq 0$, it is immediate that the evolution of $\xi_t$ is always linearly stable, and only becomes marginally unstable in the limit 
of infinitesimal adjustment rate, $\gamma \to 0$. 

This is in fact a desirable property, since the evolution equation for an economy made of a single firm is identical to that of the uniform mode. We want 
any instability to arise from the interplay between network effects and market imperfections, since the instability of a system with a single firm would be
very artificial. 

In the case where a lag is introduced and Eq. (\ref{eq:evolution3-lag}) is used instead, one finds that the uniform mode can actually become unstable if $q - q_0$ is 
sufficiently large, i.e. when the effect of the past trend on the anticipation of future prices is significantly larger than the anticipation of global 
inflation. This case corresponds to a kind of irrational optimism on the behalf of firms, who keep believing that they can sell their product at a high discounted 
price tomorrow. Although potentially interesting, we will not pursue this path further in the present work.

\subsubsection{Non-uniform modes}

\begin{figure}
\centering
\includegraphics[scale=0.6]{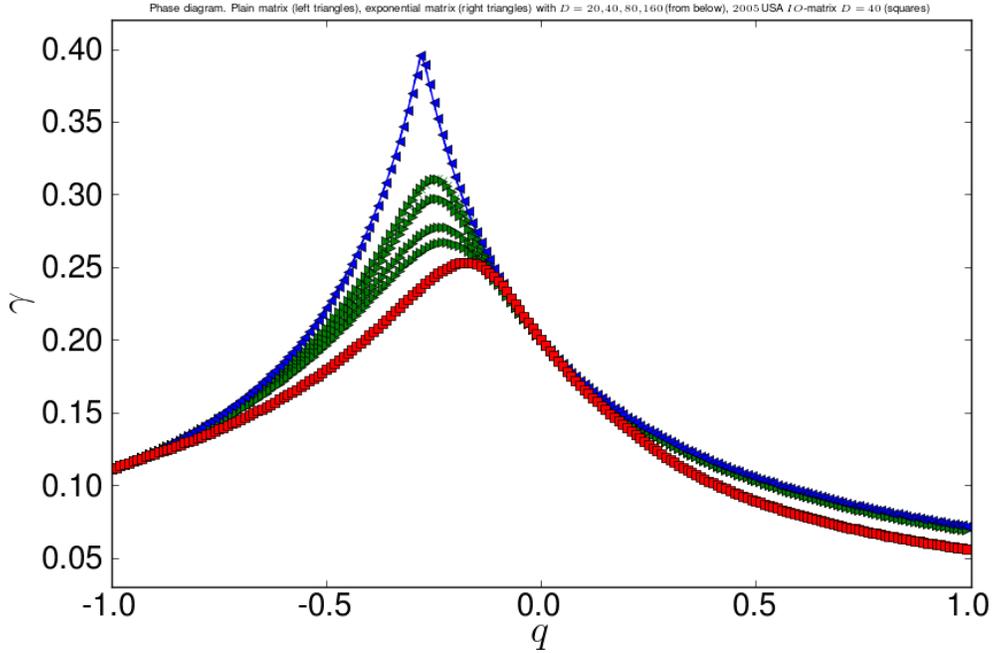}
\caption{\small{
The ``phase-diagram'' of the model in the $q, \gamma$ plane, for various type of input-output matrices, normal and non-normal. 
Below the critical line (i.e. for small enough $\gamma$), the standard rational equilibrium is dynamically stable. Above the line, the 
equilibrium is still a strict stationary solution of the dynamical model, but it is linearly unstable.  Note that the maximum value of 
$\gamma$ is reached for a slightly mean-reverting price forecast, i.e $q < 0$ but not too large.
The upper curve (left triangles) corresponds to a plain matrix of size $n=40$, the intermediate curves corresponds to random matrices
with exponentially distributed independent elements of size $n=20, 40,  80, 160$ (from bottom to top), while the lowest (most unstable) 
curve corresponds to the US input-output matrix with $n=40$.}
}
\label{fig:typ_pd}
\end{figure}

We now consider a non uniform mode $\vec V_{s} \perp \vec 1$, corresponding to another eigenvalue $s \in \mathbb{C}$ of $\mathbb{W}$, with $|s| < 1$. The evolution equation 
of the system now read (with $\epsilon_{s,t} = \vec V_{s} \cdot \vec \epsilon_t$):
\begin{align}
\mu_t - (1-a) s \pi_t &= \left(\frac{1-b}{b}\right) \xi_{t+1} - \frac{1}{b} \epsilon_{s,t} \;,\label{eq:evolution1ter}\\
(1-\gamma) (\xi_{t+1}-\xi_t) &=  \gamma\frac{b}{1-b}(\pi_t-\mu_t) -  \gamma \frac{b}{1-b}q (\pi_{t-1}-\pi_t)  \;, \label{eq:evolution2ter}\\
\xi_t + \pi_t &= (1-a)b\bar s (\mu_t + \xi_{t+1})\;\label{eq:evolutionter},
\end{align}
with $\bar s$ the complex conjugate of $s$. Eliminating $\mu_t$ between the first and third equations (and setting the noise to zero for the time being) leads to:
\begin{align}
  \pi_t &= \frac{(1-a)\bar s \xi_{t+1} - \xi_t}{1 - b(1-a)^2 |s|^2}\;,\\
  \mu_t &= \frac{\xi_t}{(1-a)b\bar s}-\xi_{t+1}+\frac{(1-a)\bar s \xi_{t+1}-\xi_t}{(1-a)b\bar s(1-b(1-a)^2|s|^2)}
\end{align}
and therefore an autonomous, second order difference equation for $\xi_t$:
\be
A_2 \xi_{t+1} + A_1 \xi_{t} + A_0 \xi_{t-1} = 0,
\ee
with $c=b(1-a) < 1$ and:
\be
A_2 = 1 - \gamma + \widehat \zeta_s (1-b-c\bar s(1+q)+c^2 |s|^2), \qquad \widehat \zeta_s = \frac{\gamma}{(1-b)(1 - b(1-a)^2 |s|^2)}
\ee
and
\be
A_1 = - \left[1 - \gamma + \widehat \zeta_s  (\bar s c(1-q)-b(1+q))\right],\qquad A_0= - q b\widehat \zeta_s.
\ee
Studying the roots of the equation $A_2 \alpha^2 + A_1 \alpha + A_0 = 0$ in full generality is 
quite involved. However, it is immediate to see that an instability with $\alpha \to 1$ cannot occur for any value of $q$, whereas $\alpha \to -1$ 
defines a certain line $\gamma_c(q)$ in the $(q,\gamma)$ plane given by (for $s$ real):
\be
\frac{1 - \gamma_c}{\gamma_c} = \frac{2b - 1 - c^2 s^2 + 2q (b + c s)}{2(1-b)(1 - b(1-a)^2 s^2)},
\ee
provided the right hand side is positive. In the limit $b \to 1$ and $s$ real, this simplifies to a more readable expression:
\be
\gamma_c \approx  \frac{2(1-(1-a) s)}{2q + 1 -(1-a) s} (1 - b),
\ee
which shows several interesting features: 
\begin{itemize}
\item a) when $b \to 1$, i.e. for constant return to scales, the system is {\it always dynamically unstable}, i.e. 
$\gamma_c = 0$; 

\item b) when $q=0$, $\gamma_c$ is independent of $a$ and $s$, and therefore of the form of the input-output network;

\item c) for a given $s$, $\gamma_c$ decreases when $q$ increases, which means that more trend following on the price (i.e. $q > 0$) destabilizes the system; 

\item d) for a given $q$, $\gamma_c$ decreases as $s$ increases. 
\end{itemize}
The numerical analysis of the roots for $a=0.5$ and $b=0.9$ leads to the phase diagram shown in Fig.~\ref{fig:typ_pd}, 
for different input-output 
matrices, including the one corresponding to the US economy. One finds that for all values of $q$, there exists a critical value of $\gamma = \gamma_c(q)$
above which the system becomes unstable as the eigenvalue $\alpha$ with the largest modulus crosses the unit circle. As anticipated from the analytical 
result above, $\gamma_c$ is approximately independent of the input-output matrix for $q=0$ and 
decreases (i.e the system becomes more unstable) when extrapolative expectations become stronger ($q \to 1$) or when mean reversion becomes strong ($q \to -1$). Interestingly however, 
one also sees that as $q$ becomes negative (i.e. the reference price is lagged further in the past, with $q \to -1$ corresponding to 
$\mathbb{E}_t(p_{t+1}) = p_{t-1}$), the instability changes nature as $\alpha$ acquires a non zero imaginary part,\footnote{When $q=-1$, $s$ real and $b \to 1$, the calculation again simplifies and leads to
a critical value $\gamma_c \approx (1-b)$ such that $\alpha = e^{\iota \theta}$, with $\cos \theta = (1 - (1-a) s)^2/2$.}
and $\gamma_c$ starts decreases again as $|q|$ increases. In other words, for a fixed value of $\gamma < \gamma_{\max}$, there is an interval $[q_-,q_+]$ within 
which the system is linearly stable, and outside which it is unstable. When $\gamma \to \gamma_{\max}$ the interval closes ($q_- \to q_+$) and for 
$\gamma > \gamma_{\max}$ the system is always unstable.  Intuitively, this means that the myopic price forecast rule prevents firms to coordinate and find the
rational equilibrium, unless firms adapt slowly to new information (i.e. if $\gamma$ is small enough).\footnote{For a similar breakdown of coordination leading 
to turbulent dynamics, see the interesting study of ``complex'' two-player games in \cite{Galla}.}
This slow adaptation allows, in a sense, the forecast errors to average out and allows the system to reach equilibrium.

Note that, quite interestingly, the structure of the input-output matrix $\mathbb{W}$ is not critical for the existence of an instability (although the
precise value of $\gamma_c$ and the detailed nature of the dynamics in the unstable phase do depend on $\mathbb{W}$). In fact, even when $\mathbb{W}$ is the 
identity matrix, the system can be unstable. The reason is that all firms are in any case globally coupled by the consumption budget of households which 
(partly) determines the demand for goods and, through the market clearing condition, the fluctuation of prices. If one visualizes the households as an extra
node in the firm network, this node is therefore connected to all firms, leading in a sense to a fragile ``star-like'' economy of the kind envisaged in \cite{Acemoglu, Acemoglu2}, 
even when $\mathbb{W}=\mathbb{I}$.

\subsection{The non-linear regime: volatility without shocks}

In the unstable phase, non-linearities start playing a role and analytical calculations become impossible, so one has to turn to numerical 
simulations of the dynamics of the 
system. Interestingly, as in many unstable dynamical systems \cite{Guck}, the non-linearities are found to stabilize the dynamics 
that becomes quasi-periodic or even chaotic, but {\it bounded}. Intuitively, the economical ingredients
of the model are indeed expected to play a stabilizing role when the system is strongly out of equilibrium: high prices strongly 
suppress demand which in turns drives prices down, etc. Let us insist once again on the fact that the standard equilibrium is still 
{\it formally} a strict solution of the dynamical equation, but has simply become an unstable (and therefore unreachable) one, leading to
either limit cycles or fully chaotic dynamics.
\begin{figure}
\centering
\includegraphics[scale=0.4]{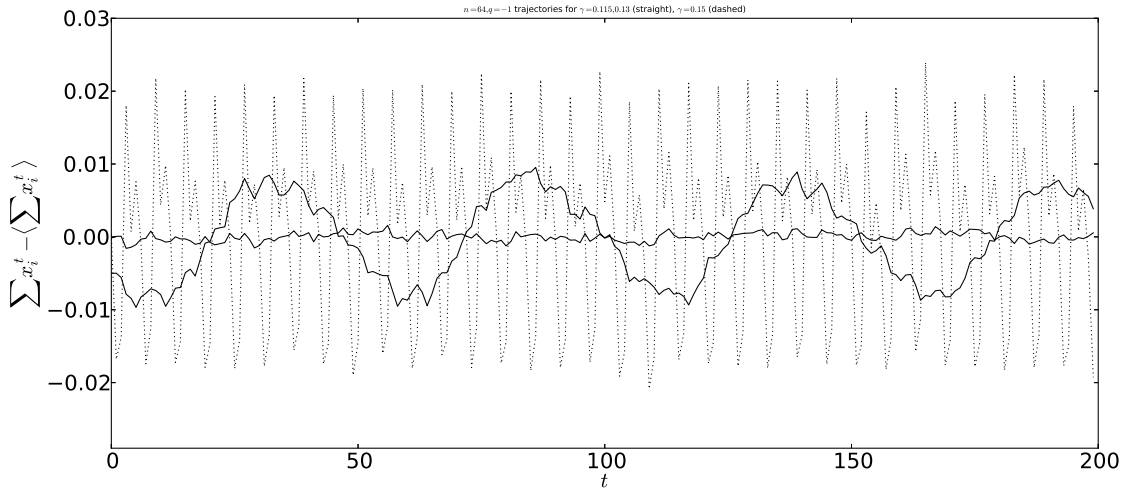}
\vspace{5mm}
\caption{\small{Three typical trajectories of aggregate output in the ``mildly'' unstable phase ($\gamma=0.115 \approx \gamma_c$ and $\gamma=0.13, 0.15 > \gamma_c$).
The standard deviation of the external shocks is very small 
($\sigma=10^{-3}$) and $q=-1$, $n=64$ in all cases. However, the aggregate volatility is considerably larger in the
unstable phase, and the dynamics displays ``business cycles''. Note also that the stationary level of aggregate output 
is above the equilibrium value in the unstable regime (see also Fig. \ref{fig:utility}).
}}
\label{fig:trac}
\end{figure}
\begin{figure}
\centering
\includegraphics[scale=0.4]{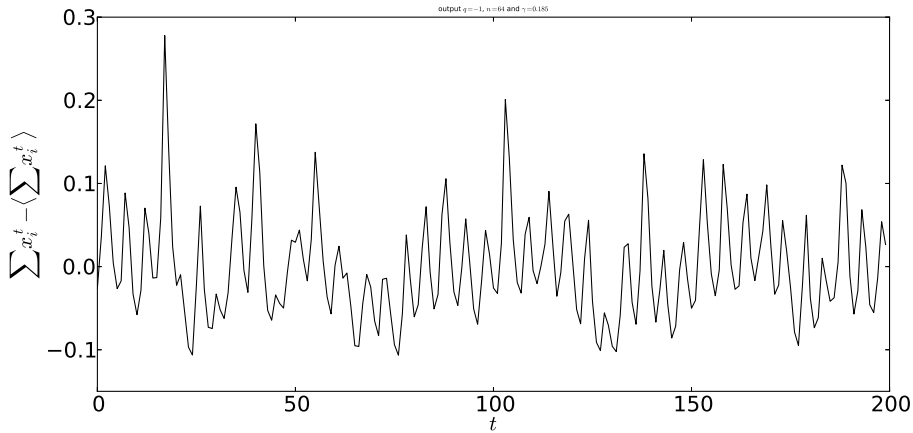}
\vspace{5mm}
\caption{\small{A typical trajectory of aggregate output in the chaotic phase ($\gamma=0.185$).
The standard deviation of the external shocks is again ($\sigma=10^{-3}$) and $q=-1$, $n=64$. Note that the 
quasi-periodic behaviour in Fig. \protect{\ref{fig:trac}} has given way to a fully irregular pattern.
}}
\label{fig:trac2}
\end{figure}
Some typical trajectories of the total output are shown in Figs.~\ref{fig:trac},\ref{fig:trac2} for a system of size $n=64$, for $q=-1$ and $\gamma=0.115$, 
$\gamma=0.13$, $\gamma=0.15$, $\gamma=0.185$. The critical point lies around $\gamma_c \approx 0.115$, but other transition points appear at higher values of
$\gamma$ as well, corresponding to different types of dynamics (quasi-periodic, chaotic), very much like physical systems undergoing transition to 
turbulence \cite{Ruelle,Guck}.  Large values of $\gamma$ lead, for large economies, to more and more chaotic dynamics, see  Fig.~\ref{fig:trac2} \cite{ustocome}. 
Let us insist that we have chosen the idiosyncratic noise $\vec \epsilon_t$ to have an extremely small variance: 
the volatility seen in Fig.~\ref{fig:trac} for $\gamma > \gamma_c$ is mostly of {\it endogenous} origin, and is a
direct consequence of the self-sustained nature of the dynamics in the unstable phase. 
Notice that $\gamma=0.13$, for example, leads to  business cycles of period $\approx 50$ time steps (12 years if the time step is interpreted as a quarter). 
Of course, the cycles generated by the dynamics are far too
regular here, one reason being that true exogenous shocks would disturb this periodicity. Similar ``business cycles'', corresponding to the limit 
cycle of a linearly 
unstable system, have been proposed in the past. One example is provided by the well known Goodwin/Lotka-Volterra oscillations, although the underlying
mechanism is completely different -- see e.g. \cite{Flaschel}.

\begin{figure}
\centering
\includegraphics[scale=0.6]{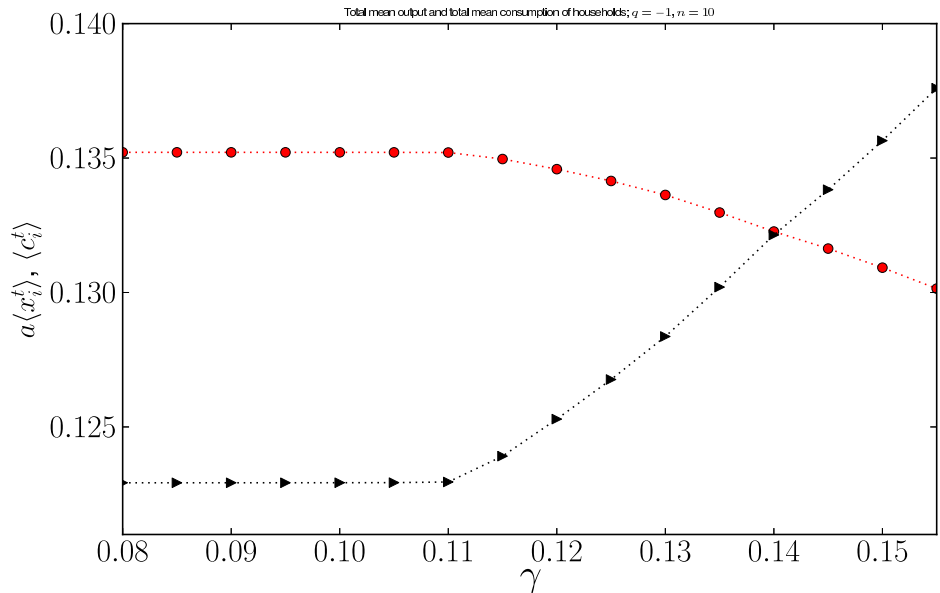}
\vspace{5mm}
\caption{\small{Average total output and total consumption of households as a function of $\gamma$, for $q=-1$. In the stable phase $\gamma < \gamma_c$, one finds -- as expected -- the theoretical equilibrium levels, 
which corresponds to an optimum in terms of household consumption. In the unstable phase $\gamma < \gamma_c$, the total output is {\it increased} compared to the equilibrium level, whereas the total household
consumption is decreased. 
}}
\label{fig:utility}
\end{figure}

Another interesting feature of the unstable phase is that the average level of the aggregate output lies \emph{above} the equilibrium level, 
whereas the average consumption of households (or their utility) decreases in the unstable phase (see Fig.~\ref{fig:utility} for more details).
The latter could have been anticipated, since the equilibrium level corresponds to an optimum welfare situation; 
the breakdown of coordination in the unstable phase leads to a reduced 
satisfaction for households but, perhaps paradoxically, to an increase of the overall output of the firms.

\begin{figure}
\centering
\includegraphics[scale=0.6]{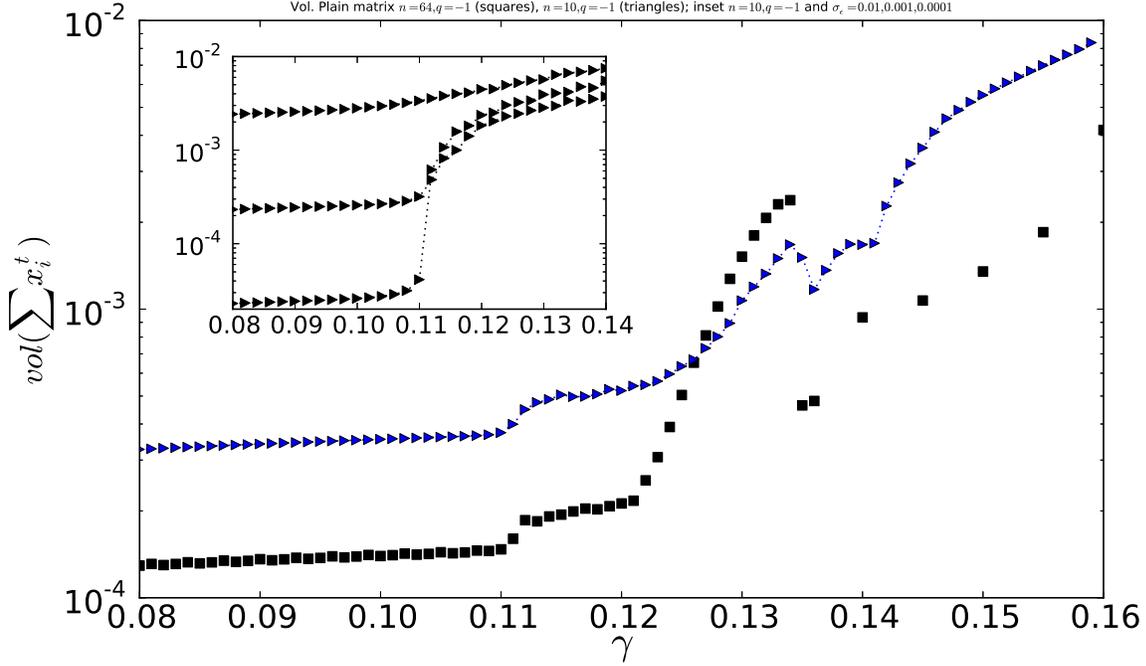}
\vspace{5mm}
\caption{\small{
Main graph: Volatility of the total output as a function of $\gamma$ for $\sigma=10^{-3}$, different sizes $n$ and a plain input-output matrix $w_{ij} \equiv 1/n$ and for $q=-1$. 
One sees that when $\gamma < \gamma_c(q=-1) \approx 0.115$, the volatility goes down with $n$, as expected for stable, balanced economies (see \protect{\cite{Horvath,Dupor,Acemoglu}}).
When $\gamma > \gamma_c$, on the other hand, the volatility remains high even as $n$ increases. The dependence of the volatility on $\gamma$ becomes highly non trivial as more modes 
become unstable as $\gamma$ increases, leading to secondary instabilities \cite{ustocome}. Inset: Volatility of the total output as a function of $\gamma$ now for a fixed value of $n=10$ but for 
$\sigma=10^{-3}, 10^{-4}$ and $10^{-5}$ (other parameters being the same than in the main graph). Now, one sees that when $\gamma < \gamma_c$, the aggregate volatility is proportional to 
that of the idiosyncratic shocks, as expected. However, in the unstable phase $\gamma > \gamma_c$ become independent of $\sigma$, even in the limit $\sigma \to 0$: one has `small shocks, but large cycles'' 
\protect{\cite{Bernanke}}.
}}
\label{fig:vol}
\end{figure}

\begin{figure}
\centering
\includegraphics[scale=0.6]{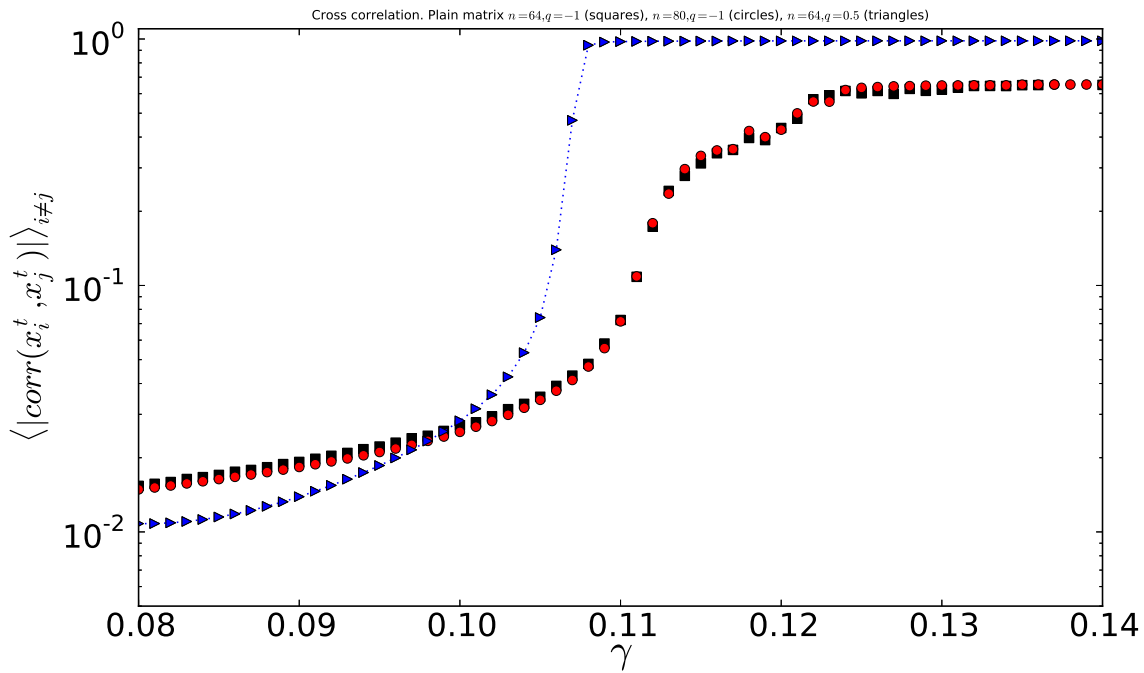}
\vspace{5mm}
\caption{\small{
Average pairwise correlation (in absolute values) of different sectors as a function of $\gamma$ for $\sigma=10^{-3}$, for $q=-1$ ($n=64$: squares and $n=80$: circles), and 
for $q=0.5$ (triangles). The input-output matrix is still the plain matrix.
One sees that when $\gamma < \gamma_c(q=-1) \approx 0.115$, the cross-correlation between sectors is small (a few $\%$) 
while it goes up to values $> 50\%$ in the unstable phase.
}}
\label{fig:cross}
\end{figure}

In order to analyse the aggregate behaviour in more detail, we plot in Fig.~\ref{fig:vol} the volatility of the
total output $\Sigma$ as a function of the adjustment parameter $\gamma$, for a given value of $q$ (here $q=-1$).\footnote{We focus here on 
total output, but have checked that other indicators, such as the consumption of households, behaves in a very similar manner.} The main graph shows $\Sigma(\gamma)$ 
for different system sizes $n$ and a given level of idiosyncratic noise $\sigma_\ell = \sigma = 10^{-3}$, whereas the inset shows  $\Sigma(\gamma)$  for a 
given $n$ and different $\sigma$'s. One clearly sees from this graph that:
\begin{itemize}
\item a) when $\gamma < \gamma_c \approx 0.115$, the volatility of the total output is 
small and goes to zero when either $\sigma \to 0$, or $n \to \infty$, as expected from the results of all previous work \cite{Acemoglu}; 
\item b) however, when $\gamma > \gamma_c$, 
the self-sustained chaotic dynamics leads to a volatility that becomes, to a good approximation, independent of $\sigma^2$ and increases quickly for all 
$n$ when $\gamma$ is increased, and {\it hence survives in the limit
of large economies and/or of vanishing idiosyncratic noise}. 
\end{itemize}
In other words, our system provides a natural framework to understand the existence of 
a business cycle in large economies, or, to paraphrase Bernanke et al. \cite{Bernanke}, the ``small shocks, large cycles puzzle''. Indeed, since 
the aggregate fluctuations are unrelated to any specific ``shock'', one cannot identify a precise
cause to the specific origin to a particular dip or peak in the total output. This is in agreement with Cochrane's conclusion in the paper cited in the 
introduction \cite{Cochrane}: 
{\it ...we [might] forever remain ignorant of the fundamental causes of economic fluctuations} -- although of course our scenario above is fundamentally different
from his.\footnote{Cochrane accounts for fluctuations by {\it ``consumption shocks," news
consumers see but we do not see. This is an attractive view, and at least explains our
persistent ignorance of the underlying shocks.} From \cite{Cochrane}.}
Another very interesting aspect of the chaotic fluctuations that the model generate is that sector fluctuations become highly correlated or anti-correlated, as announced in the
introduction, and in agreement with the conclusion of Foerster et al \cite{Foerster}. Indeed, as pointed out in the introduction we expect that close to critical point 
all sectors are driven by one instable mode and hence become perfectly correlated (or anti-correlated). In the presence of non-linear terms 
(which have not been accounted for in the introduction) several modes are driven unstable and the dynamics becomes more and more chaotic as one penetrates into the unstable 
phase.\footnote{In this sense, economic systems may become ``turbulent'', exactly as fluids do, when many modes have become unstable -- see e.g. \cite{Ruelle,Frisch}.} Hence, the cross-correlations between sectors 
remains less than $1$ but are much greater than in the stable phase. We show in Fig~\ref{fig:cross} the average absolute pairwise correlations of the fluctuations as a function of $\gamma$. Here again, we see that correlations are small in the stable phase: as emphasized in \cite{Foerster}, the correlations generated by a {\it stable} network model {\it \`a la} Long-Plosser are usually quite small, in any case much smaller than the empirically measured cross correlation. In the non-linear phase, however, the whole economy becomes driven by one (or several) unstable mode, which leads to a highly synchronised behaviour. 
In that respect, let us insist on the fact that the linear instability of the model is not that of the uniform mode, but of a non-uniform mode 
that has by definition no influence on the 
total production of the economy. But when the system is in the non-linear phase, modes become coupled and the unstable non-uniform mode plays the role of a common ``noise'' factor for the dynamical evolution of the total output. We find \cite{ustocome} that as $\gamma$ grows, the amplitude of that uniform mode grows substantially, and leads to average correlations between sectors that indeed reaches values similar to the one observed empirically ($\rho \approx 0.2$, see \cite{Foerster}).  

\section{Possible extensions and Conclusion}

The above model should really be seen as a stylized prototype, but should not to be taken too literally. 
In particular, a careful calibration of the model seems to us highly premature, since
many potentially relevant effects have been (at this stage) left out. The main reason our framework is interesting is that while remaining 
very close to the classical framework (with only two
plausible modifications: firms do not have infinitely foresight and use a myopic price forecast, 
and firms do not adjust instantaneously to the optimal production target), 
the model displays a very rich phenomenology and suggests a new way of understanding how large economies are so volatile: they are, by analogy with physical 
systems, ``turbulent''. However, many potentially important aspects of the economy
have been discarded, one of the most important being the fact that markets do not clear instantaneously, leading to stocks and/or involuntary savings. 
We have actually extended our model to account for under-production 
or surpluses, to which prices adapt more or less rapidly. Other important aspects that should be included before attempting to 
calibrate the model to real data are: savings \&
interest rates, inventories, heterogeneities of products and preferences, heterogeneous time-to-built 
(this would remove spurious effects coming from an artificial synchronisation of the
activity assumed in the above discrete time model), dynamical adaptation of the network itself (on this last aspect, see e.g. the inspiring paper \cite{Marsili}), etc. 

Still, our scenario appears to be robust and generic. Every extension that we have investigated numerically so far shows a very similar overall phenomenology: a
region of the parameter space where the rational equilibrium is stable and volatility is small, and a transition manifold beyond which the rational equilibrium cannot 
be reached dynamically and large endogenous fluctuations survive, even for large economies and vanishing idiosyncratic noise. Interestingly, we find that slow 
adjustments always help stabilizing the system: when agents attempt to reach the optimal production target too quickly, the whole economy fails to coordinate and this leads to
crises. We plan to report in full details on
these extensions, as well as on the dynamics in the chaotic phase in the near future \cite{ustocome}. 
When we are confident that the most relevant mechanisms are taken into account, a precise calibration of the enhanced model will become meaningful and in our agenda.  

\vskip 1.5cm
{\bf Acknowledgements} This work was partially financed by the CRISIS project. We want to thank all the members of CRISIS for most
useful discussions, in particular J. Batista, A. Beveratos, D. Delli Gatti, J. D. Farmer, S. Gualdi, M. Tarzia and F. Zamponi for many enlightening remarks.
We also thank A. Mandel, M. Marsili for useful inputs, in particular A. M. for pointing us to ref. \cite{Mandel}.

\end{document}